\documentclass[12pt, eqsecnum]{revtex4}
\begin{document}
\title{Finite BRST transformation and constrained systems}

\author{ Sudhaker Upadhyay\footnote {e-mail address: sudhakerupadhyay@gmail.com}}
\author{ Bhabani Prasad Mandal\footnote{e-mail address:   bhabani.mandal@gmail.com}}

\affiliation { Department of Physics,\\
Banaras Hindu University,\\
Varanasi-221005, INDIA. \\
}

\begin{abstract}
We establish the connection between the generating functional for the first-class theories 
 and the generating functional for the second-class theories 
using the finite field dependent BRST (FFBRST) 
transformation. We show this connection with the help of explicit calculations 
in two different models. The generating functional of Proca model   is obtained from the 
generating functional of Stueckelberg theory for massive spin 1 vector 
field
using FFBRST transformation.
In the other example 
we relate the generating functionals for gauge invariant and gauge variant theory for 
self-dual chiral boson. 
\end{abstract}

\maketitle

\section{Introduction}
\label{intro}
The Dirac's method for constraints analysis has been  used in great  extent in the 
Hamiltonian formalism  to quantize the system with second-class constraints \cite{dir}. 
However, the Dirac brackets which are the main ingredient in such formulation  are 
generally field 
dependent and nonlocal. Moreover, these lead  to a serious ordering problem between the
 field operators. 
These are unfavorable circumstances for finding canonically conjugate dynamical variables. On
the other hand, the quantization of the system with  first-class constraints  \cite{frad} 
has been well appreciated in a gauge invariant manner preserving BRST
symmetry \cite{brst,c,ku}. 
The system with second-class constraints can be quantized by converting these to a
first-class  theory in an extended phase space  \cite{frad,brst,c,ku,ste}. 
This procedure has
been  considered extensively by Batalin, Fradkin and Tyutin \cite{bf,bt} 
and has been applied to various models \cite{fik,kkk,fs,wz}.

The Proca model in (1+3) dimensions (4D) 
for massive spin 1 vector field  is a system with second-class constraint as  
the gauge symmetry is broken by the mass term of the theory \cite{pro,glu,lah,lah1,lah2}. However, 
Stueckelberg  converted 
this theory to a first-class theory  by introducing  a scalar 
field \cite{stuc1,stuc2,stuc3}. Such a gauge invariant description for massive spin 1
field has many application in gauge field theories as well as in string theories \cite
{mr,r,al}.

The gauge variant model  for single self-dual 
chiral boson in (1+1) dimensions (2D) is another well known example of second-class 
theory \cite{kk,pp,fj,cg,ggk,vo}. This model can be made gauge invariant by adding 
Wess-Zumino (WZ) term  and has been studied using  Batalin, Fradkin and Vilkovisky (BFV) 
formulation \cite{sub,sud}. Such a model is very useful in the study of certain string theoretic 
models \cite{ms} and plays a crucial role in the study of quantum Hall effect \cite{wen}. 

In this work we show that the theories with first-class constraints can be related to the 
theories with second-class constraint through FFBRST transformation 
introduced by Joglekar and Mandal \cite{sdj}. FFBRST transformation is a generalized 
version of usual BRST transformation where the anticommuting parameter is finite and
field dependent. Such generalized BRST transformation is nilpotent and 
also the symmetry of the effective 
action. However being finite in nature such a transformation does not leave the
path integral measure in the expression of generating functional invariant.
 Usefulness of 
such transformation  
is established through its wide applications in different field theoretic models 
\cite{sdj,rb,sdj5,susk,subp1,subp,ssb,um,um1}.
The finite field dependent anti-BRST (FF-anti-BRST) transformation 
also plays the exactly similar role as FFBRST transformation \cite{susk,subp}.  Here we show that the generating 
functional of Stueckelberg theory for massive spin 1 vector field can be related to the 
generating functional of Proca model for the same theory  through FFBRST and FF-anti-BRST transformations. 
Similar relationship is also established between the gauge invariant and gauge variant 
models for single self-dual chiral boson  through FFBRST and FF-anti-BRST formalism. The complicacy arises due to 
the nonlocal and field dependent Dirac brackets in the quantization of second-class theories can thus be avoided
by using FFBRST/FF-anti-BRST transformations which relate the Green's functions of second-class theories 
to the first-class theories.   

Here is the plan of the paper.  In Sec.II we outline the idea of the FFBRST 
transformation. We  review briefly the
theories endowed with second-class and first-class constraints for massive spin 1 field and 
self-dual chiral boson in Sec. III. The connection between first-class and
 second-class theories through FFBRST and FF-anti-BRST transformations is established in Sec. IV and Sec. V 
respectively. 
 Sec. VI is 
reserved for discussion  and conclusions.

 \section{FFBRST transformation}
In this section, we start with the 
 properties of the usual BRST transformations which do not depend on 
whether the infinitesimal BRST parameter $\Lambda$  is (i) finite or infinitesimal, (ii) 
field dependent or 
not, as long 
as it is anticommuting and space-time independent. These observations give us freedom to 
generalize the BRST transformations by making the parameter $\Lambda$, finite and field 
dependent without affecting its properties. To generalize  the BRST transformations we start 
with the making of infinitesimal BRST parameter field dependent. For the 
sake of convenience we further introduce a numerical  parameter $\kappa\ 
(0\leq \kappa\leq 1)$ and make all the fields, $\phi(x,\kappa)$, $\kappa$ dependent in such 
 a manner that $\phi(x,\kappa =0)=\phi(x)$ and $\phi(x,\kappa 
=1)=\phi^\prime(x)$, the transformed field.

The usual infinitesimal BRST transformations, thus can be written generically as 
\begin{equation}
{d\phi(x,\kappa)}=\delta_{b} [\phi (x,\kappa ) ]\Theta^\prime [\phi (x,\kappa ) ]{d\kappa},
\label{diff}
\end{equation}
where the $\Theta^\prime [\phi (x,\kappa ) ]{d\kappa}$ is the infinitesimal but field 
dependent parameter.
The finite field dependent parameter then can be 
constructed by integrating such infinitesimal  transformations from $\kappa =0$ 
to $\kappa= 1$, such that
\begin{equation}
\phi^\prime (x) =\phi(x)+\delta_b[\phi(x) ]\Theta[\phi(x) ],
\label{kdep}
\end{equation}
where 
\begin{equation}
\Theta[\phi(x)]=\int_0^1 d\kappa^\prime\Theta^\prime [\phi(x,\kappa^\prime)].\label{thet}
\end{equation}
Such FFBRST transformations are nilpotent and symmetry of the effective action. However, 
the path integral measure in the definition of generating functional is not invariant under 
such transformations as the BRST parameter is finite.
The Jacobian of the path integral measure for such transformations can be evaluated for some 
particular choice of the finite field dependent parameter, $\Theta[\phi(x)]$, as
\begin{eqnarray}
D\phi  =J(
\kappa)\ D\phi(\kappa) =J(\kappa+d\kappa)\ D\phi(\kappa+d\kappa).
\end{eqnarray}
Now the transformation from $\phi(\kappa)$ to $\phi(\kappa +d\kappa)$ is infinitesimal 
in nature. Thus
\begin{equation}
\frac{J(\kappa)}{J(\kappa +d\kappa)}=\Sigma_\phi\pm \frac{\delta\phi(x, \kappa)}{
\delta\phi(x, \kappa+d\kappa)},
\end{equation}
where $\Sigma_\phi $ sums over all fields involved in the measure and  $\pm$ 
sign refers to whether $\phi$ is a bosonic or a fermionic field.
Using the above expression we calculate the infinitesimal change in the $J(\kappa)$  as
\begin{equation}
\frac{1}{J}\frac{dJ}{d\kappa}=-\int d^4x\left [\pm \delta_b  \phi (x,\kappa )\frac{
\partial\Theta^\prime [\phi (x,\kappa )]}{\partial\phi (x,\kappa )}\right ],\label{jac}
\end{equation}

The Jacobian, $J(\kappa )$ can be replaced (within the functional integral) as
\begin{equation}
J(\kappa )\rightarrow \exp[iS_1[\phi(x,\kappa) ]],
\end{equation}
 iff the following condition is satisfied \cite{sdj}
\begin{equation}
\int {{D}}\phi (x) \;  \left [ \frac{1}{J}\frac{dJ}{d\kappa}-i\frac
{dS_1[\varphi (x,\kappa )]}{d\kappa}\right ]\exp{[i(S_{eff}+S_1)]}=0, \label{mcond}
\end{equation}
where $ S_1[\phi ]$ is local functional of fields.

By choosing appropriate $\Theta$, we can make the transformed generating 
functional $Z'\equiv \int D\phi\ e^{iS_{eff}+S_1}$ under FFBRST transformation as another 
effective theory. 
\section{The theories with constraints: Examples }
In this section, we briefly outline the essential features of second-class and
first-class theories. In particular we discuss the Proca theory for 
massive spin 1 vector field theory and gauge 
variant theory for self-dual chiral boson,  which 
are second-class theories. Corresponding first-class theories i.e. the Stueckelberg 
theory for massive spin 1 vector fields  and gauge 
invariant theory for self-dual chiral boson are also outlined in this section.
\subsection{ Theory for massive spin 1 vector field }
\subsubsection{Proca model}
We start with the action for a massive charge neutral spin 1
vector field $A_\mu$ in 4D  
\begin{equation}
S_P=\int d^4x\ {\cal L}_P,
\end{equation}
where the Lagrangian density is given as
\begin{equation}
{\cal {L}}_{P}=-\frac{1}{4} {F_{\mu\nu}}F^{\mu\nu} +\frac{M^2}{2}{A_\mu}A^\mu.\label{lag}
\end{equation}
The field strength tensor is defined as 
$
F_{\mu\nu}=\partial_\mu A_\nu -\partial_\nu A_\mu.
$
We choose the convention $g^{\mu\nu}=$ diagonal $(1,-1,-1,-1)$
 in this case, where $\mu, \nu =0, 1, 
2, 3$.
The canonically conjugate momenta for $A_\mu$ field is 
\begin{equation}
\Pi^\mu=\frac{\partial {\cal L}}{\partial \dot{A_\mu}}=F^{\mu 0}.
\end{equation}
This implies 
that the primary constraint of the theory is
\begin{equation}
\Omega_1\equiv \Pi^0 =0.
\end{equation}
The Hamiltonian density of the theory is given by
\begin{eqnarray}
{\cal H}=\Pi_\mu\dot{A^\mu} -{\cal L} =\Pi_i\partial^i A^0-\frac{1}{2}\Pi_i^2+\frac{1}{2}F_{ij}F^{ij}-\frac
{1}{2}M^2A_\mu A^\mu.
\end{eqnarray}
The time evolution for the dynamical variable $\Pi^0$ can be written as
\begin{equation}
\dot{\Pi^0}=[\Pi^0, { H}], \label{pc}
\end{equation}
where the Hamiltonian $H=\int d^3 x\ {\cal H}$.
The constraints of the theory should be invariant under time evolution and using
(\ref{pc}) we obtain the secondary constraint
\begin{equation}
\Omega_2\equiv \partial_i\Pi^i +M^2 A^0=0.
\end{equation}
Constraint $\Omega_2$ contains $A^0$ which implies that
$
[\Omega_1, \Omega_2 ]\not=0.
$
Hence, the Proca theory for massive spin 1 vector field is endowed with second-class 
constraint.

The propagator for this
theory can be written  in a simple
manner
\begin{equation}
iG_{\mu\nu}(p)=-\frac{i}{p^2-M^2}\left(\eta_{\mu\nu}-\frac{p_\mu p_\nu}{M^2}\right).
\end{equation}
Note the propagator in this theory does not fall  rapidly    
for large values of the momenta. This leads to difficulties in establishing
renormalizability of the (interacting) Proca theory for massive
photons.
Hence the limit $M \rightarrow  0$ of the Proca theory is clearly difficult to perceive.

The generating functional for the Proca theory is defined as 
\begin{equation}
Z_{P}\equiv \int DA_\mu\ e^{iS_{P}}.
\end{equation}
\subsubsection{Stueckelberg theory}
To remove the difficulties in Proca model, Stueckelberg considered the following
generalized Lagrangian density
 \begin{equation} 
S_{ST}=\int d^4x \left[-\frac{1}{4} {F_{\mu\nu}}F^{\mu\nu} +\frac{M^2}{2}\left(A_\mu -
\frac{1}{M}
\partial_\mu B\right)^2
\right],
\end{equation}
by introducing a real scalar field $B$.

This action is invariant under the following gauge transformation
\begin{eqnarray}
A_\mu (x)\rightarrow A_\mu^\prime (x)&=& A_\mu (x)+\partial_\mu \lambda(x),\\
B(x)\rightarrow B^\prime (x)&=&B(x)+M\lambda (x),
\end{eqnarray}
where $\lambda$ is gauge parameter. For the quantization of such theory one has to choose 
a gauge condition. By choosing the 't Hooft gauge condition,
${\cal L}_{gf}=-\frac{1}{2\chi}(\partial^\mu A_\mu +\chi MB)^2 $  where $\chi$ is any 
arbitrary gauge parameter, it is easy to see that 
 the propagators 
are well behaved at high momentum. As a result, there is no difficulty
in establishing renormalizability for such theory.
Now we turn to the BRST symmetry for the Stueckelberg theory. Introducing a 
ghost $(\omega)$ and antighost fields $(\omega^\star )$ the effective Stueckelberg action 
can be written as
\begin{eqnarray}
S_{ST }&=&\int d^4x \left[-\frac{1}{4} {F_{\mu\nu}}F^{\mu\nu} +\frac{M^2}{2}\left(A_\mu -
\frac{1}{M} \partial_\mu B\right)^2\right.\nonumber\\
&-&\left.\frac{1}{2\chi}(\partial^\mu A_\mu +\chi MB)^2 -\omega^\star
 (\partial^2  +
\chi M^2)\omega\right].\label{act}
\end{eqnarray}
 This
action is invariant under following on-shell BRST transformation
\begin{eqnarray}
\delta_b  A_\mu &=&\partial_\mu\omega\ \ \Lambda,\ \
\delta_b  B  =  M\omega\ \ \Lambda,\nonumber\\
\delta_b  \omega &=& 0,\ \
\delta_b \omega^\star = -\frac{1}{\chi}(\partial_\mu A^\mu +\chi MB)\ \ \Lambda,
\label{sym}
\end{eqnarray}
where $\Lambda$ is infinitesimal, anticommuting and global 
parameter.
The generating functional for Stueckelberg theory is defined as
\begin{equation}
Z_{ST }\equiv \int D\phi\ e^{iS_{ST }[\phi]},\label{zfun}
\end{equation} 
where $\phi$ is the generic notation for all fields involved in the 
theory. All the Green's functions in this theory can be obtain from $Z_{ST }$.
\subsection{Theory for self-dual chiral boson}
Self-dual chiral boson can be described by gauge variant as well as gauge  invariant 
model. The purpose of this section is to introduce such models for self-dual chiral boson. 
\subsubsection{Gauge  variant theory for self-dual chiral boson}
We start with the gauge  variant model \cite{pp} in 2D for
 single self-dual chiral boson.
The effective action for such a theory is given as 
\begin{equation} 
S_{CB}=\int d^2x\ {\cal L}_{CB} =\int d^2x\left[\frac{1}{2}\dot\varphi^2 -\frac{1}{2}{
\varphi' }^2 +\lambda ( \dot\varphi -\varphi' )\right],
\label{chi}
\end{equation} 
where over dot and prime 
denote time and space derivatives respectively and $\lambda$ is Lagrange multiplier.
The field $\varphi$ satisfies the self-duality condition $\dot\varphi =\varphi'$ 
in this case.  
We choose the Lorentz metric $g^{\mu\nu}=(1,-1)$ with  $\mu, \nu =0, 1$. The associated 
momenta for the field  $\varphi$ and Lagrange multiplier are calculated as
\begin{eqnarray}
\pi_\varphi  = \frac{\partial {\cal L}_{CB}}{\partial \dot \varphi}=\dot\varphi +
\lambda,\ \
\pi_\lambda  = \frac{\partial {\cal L}_{CB}}{\partial \dot \lambda}=0,
\end{eqnarray}
which show that the model has following primary constraint
$
\Omega_1\equiv \pi_\lambda \approx 0.
$
The   Hamiltonian density corresponding to the above Lagrangian density
 ${\cal L}_{CB}$ 
in Eq. (\ref{chi}) is
\begin{eqnarray}
{\cal H}_{CB}  = \pi_\varphi\dot\phi +\pi_\lambda\dot\lambda -{\cal L}_{CB}
 =  \frac{1}{2}(\pi_\varphi -\lambda )^2 +\frac{1}{2}\varphi'^2 +\lambda \varphi'.
\end{eqnarray}
Further we can write the total Hamiltonian density corresponding to ${\cal L}_{CB}$ by 
introducing  Lagrange multiplier field $\eta$ for the primary constraint $\Omega_1$
 as
\begin{eqnarray}
{\cal H}_{CB}^T &=& \frac{1}{2}(\pi_\varphi -\lambda )^2 +\frac{1}{2}\varphi'^2 +\lambda 
\varphi' +
\eta\Omega_1,\nonumber\\
&=&\frac{1}{2}(\pi_\varphi -\lambda )^2 +\frac{1}{2}\varphi'^2 +\lambda 
\varphi' +
\eta\pi_\lambda.
\end{eqnarray}
Following the Dirac's prescription \cite{dir}, we obtain the secondary constraint in this 
case as
\begin{eqnarray}
\Omega_2 & \equiv & \dot \pi_\lambda =[ \pi_\lambda, {\cal H}_{CB}] 
 = \pi_\varphi -
\lambda-\varphi'\approx 0.
\end{eqnarray}
The constraints $\Omega_1$ and $\Omega_2$ are of second-class as 
$[\Omega_1, \Omega_2] \not=0 $.
 This is an essential feature of a gauge variant theory.

 This model is quantized by establishing the following commutation relations \cite{pp}
\begin{eqnarray}
[\varphi(x), \pi_\varphi (y)]& =&[\varphi(x), \lambda(y) ]=+i\delta (x-y),\\
2[\lambda(x), \pi_\varphi(y)] &=&[\lambda(x), \lambda(y) ]=-2i\delta' (x-y),
\end{eqnarray}
where prime denotes the space derivative.
The rest of the commutator vanishes. 

The generating functional for gauge  variant theory for self-dual chiral boson is 
defined as
\begin{equation}
Z_{CB} =\int D\phi\ e^{iS_{CB}},
\end{equation}
where $D\phi$ is the path integral measure and $S_{CB}$ is the effective action
for self-dual chiral boson. 
\subsubsection{Gauge invariant theory for self-dual chiral boson}
To construct a gauge invariant theory corresponding to the gauge non-invariant model
for chiral bosons, one generally introduces the WZ  term  in the 
Lagrangian density
${\cal L}_{CB}$.
For this purpose we need to enlarge the Hilbert space of the theory by introducing a new 
quantum field 
$\vartheta$, called as WZ field, through the redefinition of
 fields $\varphi$ and $\lambda$  as follows \cite{wz}
$
\varphi\rightarrow \varphi -\vartheta, \ \ \lambda\rightarrow \lambda +\dot\vartheta.
$

 With these  redefinition of fields the modified Lagrangian density becomes
\begin{eqnarray} 
{\cal L}_{CB}^I={\cal L}_{CB} +{\cal L}_{CB}^{WZ},\label{wz}
\end{eqnarray}
where the WZ term
\begin{equation}
{\cal L}_{CB}^{WZ}=-\frac{1}{2}\dot\vartheta^2 -\frac{1}{2}{\vartheta'}^2  +
\varphi'\vartheta' +\dot\vartheta\vartheta' -\dot\vartheta\varphi' 
-\lambda (\dot\vartheta -\vartheta').
\end{equation}
The above Lagrangian density in Eq. (\ref{wz})  is invariant under 
time-dependent chiral gauge transformation:
\begin{eqnarray}
\delta\varphi &=&\mu (x, t),\ \ \delta\vartheta =\mu (x, t),\ \ \delta\lambda =-\dot \mu (x, 
t),
\nonumber\\
\delta\pi_\varphi &=&0,\ \ \delta\pi_\vartheta =0,\ \ \delta p_\lambda =0,
\end{eqnarray} 
where $\mu(x, t)$ is an arbitrary function of the space and time.

The  BRST invariant effective theory for self-dual chiral boson \cite{sud}
can be written as
\begin{eqnarray}
S_{CB}^{II}&=&\int d^2x\ {\cal L}_{CB}^{II}, \label{lagc}\\{\mbox{where}}\ \
{\cal L}_{CB}^{II}&=&\frac{1}{2}\dot\varphi^2 -\frac{1}{2}{\varphi'}^2 +\lambda (\dot\varphi 
-\varphi' ) -\frac{1}{2}\dot\vartheta^2 -\frac{1}{2}{\vartheta'}^2 +
\varphi'\vartheta' \nonumber\\
&+&\dot\vartheta\vartheta' -\dot\vartheta\varphi' 
-\lambda (\dot\vartheta -\vartheta')
-\frac{1}{2}(\dot \lambda -\varphi -\vartheta )^2
+\dot{\bar c}\dot c 
-2\bar c c. \label{actcb}
\end{eqnarray} 
 $c$ and  $\bar c$ are ghost and antighost fields respectively.
Corresponding generating functional for gauge invariant theory for self-dual chiral boson is 
given as
\begin{equation}
Z_{CB}^{II} =\int D\phi\ e^{iS_{CB}^{II}},
\end{equation}
where $\phi$ is generic notation for all fields involved in the effective action.
The effective action $ S_{CB}^{II}$ and the generating functional $ Z_{CB}^{II}$ are 
invariant under the
 following nilpotent BRST transformation  
\begin{eqnarray}
\delta_{b}\varphi &=&  c\ \Lambda,\ \ \ \delta_{b}\lambda =-\dot { c
}\ \Lambda, \ \ \ \delta_{b}\vartheta = c\ \Lambda,\nonumber\\
\delta_{b}\bar c &=& -(\dot \lambda -\varphi -\vartheta )\ \Lambda,\ \ \ \delta_{b}  c =0,
\end{eqnarray}
where $\Lambda$  is infinitesimal and anticommuting  BRST parameter.          
\section{Relating the first-class and second-class theories through FFBRST formulation: Examples}
In this section, we consider two examples to show the connection between the generating
 functionals for theories with first-class  and second-class constraints. Firstly 
 we show the connection between Stueckelberg theory and Proca theory for massive
vector fields. In  the second example
 we link  the gauge invariant and gauge variant theory for self-dual chiral 
 boson.
\subsection{Relating Stueckelberg and Proca theories } 
We start with the linearize form of the 
 Stueckelberg effective action (\ref{act}) by introducing a Nakanishi-Lautrup type auxiliary
field ${\cal B}$ as  
\begin{eqnarray}
S_{ST}&=&\int d^4x \left[-\frac{1}{4}F_{\mu\nu}F^{\mu\nu} +\frac{1}{2}M^2\left(A_\mu -
\frac{1}{M}\partial_\mu B\right)^2 \right.\nonumber\\
 &+&\left.\frac{\chi}{2}{\cal B}^2-{\cal B}(\partial_\mu A^\mu 
+\chi MB) 
-\omega^\star (\partial^2 +\chi M^2)\omega\right],\label{s}
\end{eqnarray}
which is invariant under the following off-shell nilpotent BRST transformation
\begin{eqnarray}
\delta_b A_\mu =\partial_\mu\omega\ \ \Lambda,\ \
\delta_b B  =  M\omega\ \ \Lambda,\ \
\delta_b \omega  =  0,\ \
\delta_b \omega^\star  =  {\cal B}\ \ \Lambda,\ \
\delta_b {\cal B} =0.
\end{eqnarray}
The FFBRST transformation corresponding to the above BRST transformation is constructed as,
\begin{eqnarray}
\delta_b A_\mu =\partial_\mu\omega\ \Theta_1[\phi ], \ \
\delta_b B = M\omega\ \Theta_1[\phi ],\ \
\delta_b \omega =0,\ \
\delta_b \omega^\star ={\cal B}\  \Theta_1[\phi ],\ \
\delta_b {\cal B}=0,\label{fin}
\end{eqnarray}
where $\Theta_1$ is an arbitrary finite field dependent parameter but still anticommuting 
in nature.
To establish the connection we choose a finite field dependent parameter $\Theta_1$ 
obtainable from   
\begin{equation}
\Theta_1^\prime =i\gamma\int d^4x\left[\omega^\star \left(\chi MB-\frac{\chi}{2}{\cal B}+
\partial_\mu A^\mu\right)\right],
\end{equation}
via Eq. (\ref{thet}), where $\gamma$ is an arbitrary parameter.

Using Eq. (\ref{jac}) the infinitesimal change in nontrivial Jacobian can be calculated 
for this finite field dependent parameter as
\begin{eqnarray}
\frac{1}{J}\frac{dJ}{d\kappa}&=& i\gamma\int d^4x \left[ {\cal B}
\left(\chi MB-\frac{\chi}{2}{\cal B}+
\partial_\mu A^\mu\right) 
\right],
\end{eqnarray}
where the equation of motion for antighost field,
$(\partial^2 +\chi M^2)\omega =0$, has been used.
 
We now make the following ansatz for  $S_1$ as
\begin{eqnarray}
S_1=\int d^4x [\xi_1 (\kappa )\ {\cal B}^2+\xi_2 (\kappa )\ {\cal B} \partial_\mu A^
\mu+\xi_3
 (\kappa )\ \chi M{\cal B}B  ],
\end{eqnarray}
where $\xi_i,\ (i=1, 2, 3)$ are arbitrary $\kappa$ dependent parameter and
satisfy following initial conditions $\xi_i (\kappa =0)=0 $. 
Now, using the relation in Eq. (\ref{diff}) we calculate $\frac{dS_1}{d\kappa}$ as 
\begin{eqnarray}
\frac{dS_1}{d\kappa}=\int d^4x \left [ {\cal B}^2 \xi_1^\prime 
 + {\cal B}\partial_\mu A^\mu 
\xi_2^\prime  + \chi M {\cal B}B \xi_3^\prime  
 \right],
\end{eqnarray}
where prime denote the differentiation  with respect to $\kappa$.
The Jacobian contribution can be written  as $e^{S_1}$ if the  essential condition in 
Eq. (\ref{mcond}) is satisfied. This leads to
\begin{eqnarray}
\int d^4x e^{i(S_{ST}+S_1)}\left [i{\cal B}^2(\xi_1^\prime +\gamma \frac{\chi}{2}
 )
+i{\cal B}\partial_\mu A^\mu(\xi_2^\prime -\gamma )
+i\chi M {\cal B}B(\xi_3^\prime 
-\gamma )
\right]
=0.\label{mcond1}
\end{eqnarray}
Equating the coefficient of terms $ i{\cal B}^2, i{\cal B}\partial_\mu A^\mu, $ and $
i\chi M {\cal B}B$
from both sides
of above condition, 
we get following  differential equations 
\begin{eqnarray}
\xi_1^\prime +\gamma \frac{\chi}{2}&=&0, \ \
\xi_2^\prime -\gamma  = 0,\ \
\xi_3^\prime -\gamma  = 0.
\end{eqnarray}
To obtain the solution of the above equations we put $\gamma =1$ without any loose 
of generality. The solutions satisfying initial conditions are given as
\begin{equation}
\xi_1=-\frac{\chi}{2}\kappa,\ \xi_2=\kappa,\ \xi_3=\kappa.
\end{equation}
The transformed action can be obtained by adding $S_1(\kappa=1)$ to $S_{ST}$ as 
\begin{eqnarray} 
S_{ST}+S_1&=&\int d^4x \left[-\frac{1}{4}F_{\mu \nu } F^{\mu \nu }+\frac{1}{2}M^2
\left(A_\mu  
-\frac{1}{M}\partial_\mu B\right)^2 -\omega^\star (\partial^2 +\chi M^2)\omega\right].
\end{eqnarray}
Now the   generating functional under FFBRST transforms as
 \begin{eqnarray}
 Z'= \int DA_\mu DBD\omega D\omega^\star e^{i(S_{ST }+S_1)}.
 \end{eqnarray}
To remove the divergence due to gauge volume in the above expression we integrate
  over the  $B, \omega,$ and $ \omega^\star$ fields  
which reduces it to the generating functional for Proca model upto some normalization constant as follows 
 \begin{eqnarray}
 Z'= \int DA_\mu \ e^{i\int d^4 x\left[ -\frac{1}{4} {F_{\mu\nu}}F^{\mu\nu} +\frac{M^2}{2}{A_\mu}A^\mu \right]}= Z_P.
 \end{eqnarray}
 Here we would like to point out that this action is divergence free as mass term breaks the gauge symmetry.
Therefore,
\begin{equation}
Z_{ST}\left(\int D\phi\ e^{iS_{ST }}\right)\stackrel{FFBRST}{---\longrightarrow} Z_P
\left(\int DA_\mu\ e^{iS_P}\right).
\end{equation}
Thus by constructing appropriate finite field dependent 
parameter (given in Eq. (\ref{fin})) we have shown that the generating functional for Stueckelberg 
theory is connected to the 
generating functional  for Proca theory through FFBRST transformation. This
indicates that the Green's functions in these two theories are related through FFBRST 
formulation.
\subsection{Relating the gauge invariant and variant theory for chiral boson  }
To see the connection between the gauge invariant and variant  theories for chiral
 boson, we start with the effective action  
for the  gauge invariant self-dual
 chiral boson theory as  
\begin{eqnarray} 
S_{CB}^{II}&=&\int d^2x
\left[\frac{1}{2}\dot\varphi^2 -\frac{1}{2}{\varphi'}^2 +\lambda (\dot\varphi 
-\varphi' ) -\frac{1}{2}\dot\vartheta^2 
-\frac{1}{2}{\vartheta'}^2 
+\varphi'\vartheta' \right.\nonumber\\
&+&\left.\dot\vartheta\vartheta' -\dot\vartheta\varphi' 
-\lambda (\dot\vartheta -\vartheta')
+\frac{1}{2}{\cal B}^2  
 + {\cal B}(\dot \lambda -\varphi -\vartheta )
+\dot{\bar c}\dot c 
-2\bar c c\right],\label{action}
\end{eqnarray}  
where we have linearized the gauge fixing part of the effective action by introducing the
extra auxiliary field ${\cal B}$.
 This effective action is invariant under following infinitesimal BRST transformation
 \begin{eqnarray}
\delta_{b}\varphi &=&  c\ \Lambda,\ \ \ \delta_{b}\lambda =-\dot { c
}\ \Lambda, \ \ \ \delta_{b}\vartheta = c\ \Lambda,\nonumber\\
\delta_{b}\bar c &=& {\cal B}\ \Lambda,\ \ \ \delta_{b} {\cal B} =0, \ \ \ \delta_{b}  c =0.
\end{eqnarray} 
Corresponding  FFBRST transformation can be written as
\begin{eqnarray}
\delta_{b}\varphi &=&  c\ \Theta_1 [\phi],\ \ \ \delta_{b}\lambda =-\dot { c
}\ \Theta_1 [\phi ], \ \ \ \delta_{b}\vartheta = c\ \Theta_1 [\phi],\nonumber\\
\delta_{b}\bar c &=& {\cal B}\ \Theta_1 [\phi],\ \ \ \delta_{b} {\cal B} =0, 
\ \ \ \delta_{b}  c =0,\label{fbrs}
\end{eqnarray} 
where $\Theta_1 [\phi ]$ is arbitrary finite field dependent parameter,
which we have to constructed.
In this case we construct the  finite field dependent BRST parameter $\Theta_1 [\phi]$
obtainable from
\begin{equation}
\Theta_1'= i\gamma\int d^2x \left[\bar c (\dot \lambda -\varphi-\vartheta+\frac{1}{2}{\cal B})
\right],
\end{equation} 
using Eq. (\ref{thet})
and demand that the corresponding BRST transformation will lead to the gauge variant theory 
 for self-dual chiral boson.
  
To justify our claim we calculate the change in Jacobian, using equation of motion 
for antighost field, as
\begin{eqnarray}
\frac{1}{J}\frac{dJ}{d\kappa}&=& i\gamma\int d^2x \left[ {\cal B}(
\dot\lambda-\varphi-\vartheta 
+\frac{1}{2}{\cal B})\right].
\end{eqnarray}
We make an ansatz for local functional $S_1$ as, 
\begin{eqnarray}
S_1&=&\int d^2x \left[\xi_1(\kappa)\ {\cal B}^2 +\xi_2 (\kappa)\ {\cal B}
(\dot\lambda-\varphi-\vartheta )
 \right].
\end{eqnarray}
The  change in $S_1$ with respect to $\kappa$ is calculated as 
\begin{eqnarray} 
\frac{dS_1}{d\kappa} = \int d^2x \left[\xi_1' \ {\cal B}^2 +\xi_2'\ {\cal B}(
\dot\lambda-\varphi-\vartheta )
\right]. 
\end{eqnarray}
Now, the necessary condition in 
Eq. (\ref{mcond}) leads to the following equation 
\begin{eqnarray} 
\int  d^2x e^{i(S_{CB}^{II}+S_1)}\left [i{\cal B}^2(\xi_1^\prime - \frac{\gamma}{2}
 )+i{\cal B}(\dot\lambda-\varphi-\vartheta )(\xi_2^\prime -\gamma )
 \right]
 =0.\label{mcond2} 
\end{eqnarray}
Equating the coefficient of terms $ i{\cal B}^2$ and $ i{\cal B}(\dot\lambda-\varphi-\vartheta )$ from both sides
of above condition, 
we get following differential equations:
\begin{eqnarray}
\xi_1^\prime - \frac{\gamma}{2}&=&0,\ \
\xi_2^\prime -\gamma  = 0.
\end{eqnarray}
The solutions of above equations 
are
$
\xi_1=-\frac{1}{2}\kappa,\ \xi_2=-\kappa,$
where we have taken the parameter $\gamma =-1$.
The transformed action is obtained by adding $S_1(\kappa=1)$ to $S_{CB}^{II}$ as
\begin{eqnarray}
S_{CB}^{II}+S_1&=&\int d^2x \left[\frac{1}{2}\dot\varphi^2 -\frac{1}{2}{\varphi'}^2 +\lambda 
(\dot\varphi -\varphi' ) -\frac{1}{2}\dot\vartheta^2\right.\nonumber\\
&-&\left.\frac{1}{2}{\vartheta'}^2 +
\varphi'\vartheta' +\dot\vartheta\vartheta' -\dot\vartheta\varphi' 
-\lambda (\dot\vartheta -\vartheta') +\dot{\bar c}\dot c 
-2\bar c c\right].
\end{eqnarray}
Now the transformed generating functional becomes
\begin{eqnarray}
Z' =\int D\varphi D\vartheta D\lambda Dc D\bar c\ e^{i(S_{CB}^{II}+S_1)}.
\end{eqnarray}
Performing integration over fields $\vartheta, c$ and $ \bar c$,
the above generating functional reduces to the generating functional for self dual chiral boson 
upto some constant as
\begin{eqnarray}
Z' =\int D\varphi  D\lambda \ e^{iS_{CB}}= Z_{CB}.
\end{eqnarray}
Therefore,
\begin{equation}
Z_{CB}^{II}\left(\int D\phi\ e^{iS_{CB}^{II}}\right)\stackrel{FFBRST}{ --\longrightarrow} 
Z_{CB}\left(\int  D\varphi  D\lambda\ e^{iS_{CB}}\right).
\end{equation} 
Thus, the generating functionals corresponding to the gauge invariant and gauge
non-invariant theory for self-dual chiral boson 
are connected through the FFBRST transformation given in Eq. (\ref{fbrs}).
 
We end up this section by making conclusion that using FFBRST formulation 
the generating functional for the  theory with second-class constraint can be 
achieved by generating functional for   theory with first-class constraint.
\section{Relating the first-class and second-class theories through FF-anti-BRST formulation: Examples}
In this section, we consider FF-anti-BRST formulation  to show the connection between the generating
 functionals for theories with first-class  and second-class constraints with same examples. 
The FF-anti-BRST transformation is also developed in  same fashion as FFBRST transformation,
the only key difference is the role of ghost fields are interchanged with anti-ghost fields 
and vice-versa. 
\subsection{Relating Stueckelberg and Proca theories} 
We start with anti-BRST symmetry transformation for effective action given in Eq. (\ref{s}), as  
\begin{eqnarray}
\delta_{ab} A_\mu &=&\partial_\mu\omega^\star\ \ \Lambda,\ \
\delta_{ab} B  =  M\omega^\star\ \ \Lambda,\ \
\delta_{ab} \omega  =  -{\cal B}\ \ \Lambda,\nonumber\\
\delta_{ab} \omega^\star &=& 0,\ \
\delta_{ab} {\cal B} = 0,
\end{eqnarray}
where $\Lambda$ is infinitesimal, anticommuting and global parameter. 
The FF-anti-BRST transformation corresponding to the above anti-BRST transformation is constructed as,
\begin{eqnarray}
\delta_{ab} A_\mu &=&\partial_\mu\omega^\star\ \ \Theta_2,\ \ 
\delta_{ab} B  =  M\omega^\star\ \ \Theta_2,\ \
\delta_{ab}\omega  = -{\cal B}\ \ \Theta_2,\nonumber\\
\delta_{ab} \omega^\star &=& 0,\ \
\delta_{ab} {\cal B} = 0,\label{fin11} 
\end{eqnarray}
where $\Theta_2$ is an arbitrary finite field dependent parameter but still anticommuting 
in nature.
To establish the connection we choose a finite field dependent parameter $\Theta_2$ 
obtainable from     
\begin{equation}
\Theta_2^\prime =-i\gamma\int d^4x\left[\omega \left(\chi MB-\frac{\chi}{2}{\cal B}+
\partial_\mu A^\mu\right)\right], 
\end{equation}
  where $\gamma$ is an arbitrary parameter.

Using Eq. (\ref{jac}) the infinitesimal change in nontrivial Jacobian can be calculated 
for this finite field dependent parameter as
\begin{eqnarray}
\frac{1}{J}\frac{dJ}{d\kappa}&=&-i\gamma\int d^4x \left[-{\cal B}
\left(\chi MB-\frac{\chi}{2}{\cal B}+
\partial_\mu A^\mu\right) 
\right].\label{j2} 
\end{eqnarray}
To Jacobian contribution can be expressed as $e^{iS_2}$. To calculate  $S_2$ we make following ansatz 
\begin{eqnarray}
S_2&=&\int d^4x [\xi_5 (\kappa ){\cal B}^2+\xi_6 (\kappa ){\cal B} \partial_\mu A^
\mu+\xi_7
 (\kappa )\chi M{\cal B}B ],
\end{eqnarray}
where $\xi_i,\ (i=5,..,7)$ are arbitrary $\kappa$ dependent parameter and
satisfy following initial conditions $\xi_i (\kappa =0)=0 $. 
Now, infinitesimal change in $S_2$ is calculated as 
\begin{eqnarray} 
\frac{dS_2}{d\kappa}=\int d^4x \left [ {\cal B}^2 \xi_5^\prime 
 + {\cal B}\partial_\mu A^\mu 
\xi_6^\prime  + \chi M {\cal B}B \xi_7^\prime 
\right],\label{s2} 
\end{eqnarray}
where prime denotes the differentiation  with respect to $\kappa$.

Putting the expressions (\ref{j2}) and (\ref{s2})  in the  essential condition given in 
Eq. (\ref{mcond}), we obtain
\begin{eqnarray}
 \int d^4x\ e^{i(S_{ST}+S_2)}\left [ {\cal B}^2(\xi_5^\prime +\gamma \frac{\chi}{2}
 )
+ {\cal B}\partial_\mu A^\mu(\xi_6^\prime -\gamma )
+ \chi M {\cal B}B(\xi_7^\prime 
-\gamma )\right]
 =0.\label{mcond3}
\end{eqnarray}  
Equating the coefficient of terms $ i{\cal B}^2, i{\cal B}\partial_\mu A^\mu,$ and $ 
i\chi M {\cal B}B$
  from both sides
of above condition, 
we get following  differential equations 
\begin{eqnarray}
\xi_5^\prime +\gamma \frac{\chi}{2}=0,\ \
\xi_6^\prime -\gamma =0,\ \ 
\xi_7^\prime -\gamma =0.  
\end{eqnarray}
 The solutions of the above differential equation for $\gamma =1$ are
$
\xi_5=-\frac{\chi}{2}\kappa,\ \xi_6=\kappa,\ \xi_7=\kappa.
$
The transformed action can be obtained by adding $S_2(\kappa=1)$ to $S_{ST}$ as 
\begin{eqnarray} 
S_{ST}+S_2&=&\int d^4x \left[-\frac{1}{4}F_{\mu \nu } F^{\mu \nu }+\frac{1}{2}M^2
\left(A_\mu  
-\frac{1}{M}\partial_\mu B\right)^2 -\omega^\star (\partial^2 +\chi M^2)\omega\right].
\end{eqnarray}
We perform integration over $B, \omega$ and $ \omega^\star$ fields to remove the divergence 
of transformed generating functional $Z'= \int DA_\mu DBD\omega D\omega^\star e^{i(S_{ST}+S_2)}$ and hence
we get the generating functional for Proca model as 
\begin{eqnarray} 
Z'= \int DA_\mu e^{i\left[  -\frac{1}{4} {F_{\mu\nu}}F^{\mu\nu} +\frac{M^2}{2}{A_\mu}A^\mu  \right]}=  Z_{P }, 
\end{eqnarray} 
which is a divergence free theory.
Therefore, 
$
Z_{ST}\stackrel{FF-anti-BRST}{-----\longrightarrow} Z_P.
$
Thus by constructing appropriate finite field dependent 
parameter (given in Eq. (\ref{fin11})) we have shown that the generating functional for Stueckelberg 
theory is related to the 
generating functional  for Proca theory through FF-anti-BRST transformation also. This
indicates that the Green's functions in these two theories are related through FFBRST and FF-anti-BRST 
transformation.
\subsection{Relating the gauge invariant and variant theory for chiral boson }
To connect the gauge invariant and variant  theories for chiral
 boson through FF-anti-BRST transformation, first of all we write the anti-BRST transformation for effective action 
 (\ref{action}) as
     \begin{eqnarray}
\delta_{ab}\varphi &=&  \bar c\ \Lambda,\ \ \ \delta_{ab}\lambda =-\dot {\bar c
}\ \Lambda, \ \ \ \delta_{ab}\vartheta = \bar c\ \Lambda,\nonumber\\
\delta_{ab} c &=& -{\cal B}\ \Lambda,\ \ \ \delta_{ab} {\cal B} =0, \ \ \ \delta_{ab} \bar  c =0.
\end{eqnarray}
Corresponding  FF-anti-BRST transformation can be written as
\begin{eqnarray}
\delta_{ab}\varphi &=& \bar  c\ \Theta_2 [\phi],\ \ \ \delta_{ab}\lambda =-\dot {\bar  c
}\ \Theta_2 [\phi ], \ \ \ \delta_{ab}\vartheta =\bar  c\ \Theta_2 [\phi],\nonumber\\
\delta_{ab}c &=& -{\cal B}\ \Theta_2 [\phi],\ \ \ \delta_{ab} {\cal B} =0, 
\ \ \ \delta_{ab} \bar  c =0,\label{fbrs1}
\end{eqnarray} 
where $\Theta_2 [\phi ]$ is arbitrary finite field dependent parameter,
which we have to construct.
In this case we construct the  finite field dependent anti-BRST parameter $\Theta_2 [\phi]$
obtainable from
\begin{equation}
\Theta_2'= -i\gamma\int d^2x \left[ c (\dot \lambda -\varphi-\vartheta+\frac{1}{2}{\cal B})
\right], 
\end{equation} 
using Eq. (\ref{thet})
and demand that the corresponding anti-BRST transformation will lead to the gauge variant theory 
 for self-dual chiral boson. 
  
We make an ansatz for local functional $S_2$ in this case as, 
\begin{eqnarray}
S_2&=&\int d^2x \left[\xi_5(\kappa)\ {\cal B}^2 +\xi_6 (\kappa)\ {\cal B}
(\dot\lambda-\varphi-\vartheta )
\right].
\end{eqnarray}
Now, the necessary condition in 
Eq. (\ref{mcond}) leads to the following equation 
\begin{eqnarray} 
\int d^2xe^{i(S_{CB}^{II}+S_2)}\left [i{\cal B}^2(\xi_5^\prime - \frac{\gamma}{2}
 )+i{\cal B}(\dot\lambda-\varphi-\vartheta )(\xi_6^\prime -\gamma )
 \right]
 =0.\label{mcond4} 
\end{eqnarray}
Equating the coefficient of different terms in the both sides
of above equation, 
we get following differential equations:
\begin{eqnarray}
\xi_5^\prime - \frac{\gamma}{2} = 0,\ \
\xi_6^\prime -\gamma  = 0.
\end{eqnarray}
The solutions of above equations 
are 
$
\xi_5=-\frac{1}{2}\kappa,\ \xi_6=-\kappa,
$
where the parameter $\gamma =-1$.
The transformed action can be obtained by adding $S_2(\kappa=1)$ to $S_{CB}^{II}$ as
\begin{eqnarray}
S_{CB}^{II}+S_2&=&\int d^2x \left[\frac{1}{2}\dot\varphi^2 -\frac{1}{2}{\varphi'}^2 +\lambda 
(\dot\varphi -\varphi' ) -\frac{1}{2}\dot\vartheta^2\right.\nonumber\\
&-&\left.\frac{1}{2}{\vartheta'}^2 +
\varphi'\vartheta' +\dot\vartheta\vartheta' -\dot\vartheta\varphi' 
-\lambda (\dot\vartheta -\vartheta')+\dot{\bar c}\dot c -2\bar c c\right].
\end{eqnarray}
After functional integration over fields $\vartheta, c$ and $\bar c$ in the expression of
transformed generating functional, we get the generating functional as 
\begin{eqnarray}
Z' =\int D\varphi D\lambda e^{i S_{CB}}=  Z_{CB}.
\end{eqnarray}
Therefore,
$Z_{CB}^{II} \stackrel{FF-anti-BRST}{-----\longrightarrow} Z_{CB}.$
Thus, the generating functionals corresponding to the gauge invariant and gauge
non-invariant theory for self-dual chiral boson 
are also connected through the FF-anti-BRST transformation given in Eq. (\ref{fbrs1}).

We end up this section by making comment that  
the generating functional for the  theory with second-class constraint can be 
obtained from generating functional for   theory with first-class constraint
using both FFBRST and FF-anti-BRST transformations.
\section{Concluding remarks} 
The Stueckelberg theory for the  massive spin  1 field and gauge invariant 
theory for self-dual chiral boson are  the  first-class  theories. On the other hand,
the Proca theory  for massive spin 1 field and gauge  variant 
theory for self-dual chiral boson are  theories with second-class constraint. 
We have shown that the generalized BRST transformation,  where the 
 BRST parameter is finite and field dependent, relates
the generating functionals of second-class theory  and first-class
theory. The path integral measure in the definition of generating functionals
 are not invariant under such FFBRST 
transformation and are responsible for such  connections. The Jacobian for 
path integral measure under such a  transformation with appropriate finite 
parameter cancels the extra parts of
the first-class theory. We have avoided the complicated calculations of nonlocal and field dependent
Dirac brackets for second-class theory in the cost of calculating the Jacobian of nontrivial 
FFBRST transformation. 
Our result is supported by two explicit examples. In the first case we have related  the 
generating functional of Stueckelberg  
theory to the generating functional of
Proca model and in the second case the generating functionals corresponding to the gauge 
invariant theory and  gauge variant theory for self-dual chiral boson 
have linked through 
FFBRST transformation with appropriate choices of finite field dependent parameter.
Same goal has been achieved by using FF-anti-BRST transformation.
These formulations can be applied to connect the generating functionals for any first-class (e.g. non-Abelian
gauge theories)
and second-class theories provided  appropriate 
finite parameters are constructed.

\section*{Acknowledgments}

We thankfully acknowledge the financial support from the Department of Science and Technology 
(DST), Government of India, under the SERC project sanction grant No. SR/S2/HEP-29/2007.

\end{document}